\documentclass[twocolumn,aps,prl,showpacs]{revtex4}
\usepackage{amsmath} 
\usepackage{amssymb} 
\usepackage{epsfig} 

\begin{document} 
\title{Quantum critical behavior of electrons at the edge of charge order} 
\author{L. Cano-Cort\'es$^1$, J. Merino$^1$ and S. Fratini$^2$} 
\affiliation{
$^1$ Departamento de F\'\i sica Te\'orica de la Materia Condensada,
Universidad Aut\'onoma de Madrid, Madrid 28049, Spain\\ 
$^2$ Institut N\'eel-CNRS and Universit\'e Joseph Fourier,Bo\^ite
Postale 166, F-38042 Grenoble Cedex 9, France}

\begin{abstract}
We consider quantum critical points (QCP) 
in which quantum fluctuations associated with charge rather than magnetic order induce
unconventional metallic 
properties. Based on finite-$T$ calculations on a two-dimensional
extended Hubbard model 
we show how the coherence scale $T^*$ characteristic of Fermi liquid behavior
of the homogeneous metal vanishes at the onset of charge order.
A strong effective mass enhancement
reminiscent of heavy fermion  
behavior indicates the possible    
destruction of quasiparticles at the
QCP. Experimental probes on quarter-filled layered organic materials are proposed for unveiling the behavior of electrons
across the quantum critical region.   
\end{abstract}
\date{\today} 
\pacs{71.10.Hf;74.40.Kb;74.70.Kn;71.10.Fd}
\maketitle

\paragraph{Introduction.}

Quantum critical points occur at zero temperature second order
phase transitions in 
which the strength of quantum fluctuations is controlled by an
external field such 
as pressure, magnetic field or chemical composition \cite{Sachdev}. In
recent years intensive studies 
have focused on itinerant electrons at the
edge of magnetic order, being the heavy fermion
materials \cite{Coleman,Si,Lohneysen} prototypical examples.  
Much less explored but equally interesting are QCP arising from
tuning the  electrons close to a charge ordering instability.
This situation is realized in the quarter-filled families 
of layered organic superconductors \cite{Ishiguro} of the 
$\alpha$, $\beta''$ and $\theta$-(BEDT-TTF)$_2$X types.
Large electron effective mass enhancements and non-Fermi liquid 
metallicity at finite-$T$ are observed in 
(MeDH-TTP)$_2$AsF$_6$ and $\kappa$-(DHDA-TTP)$_2$SbF$_6$
above a critical pressure at which the charge order found at ambient pressure melts\cite{Yasuzuka,Weng}. Such heavy fermion behavior may appear puzzling,
considering the different $\pi$-orbitals of the organics as compared to 
the f-orbitals in the rare earths, but can find a natural explanation
based on the universal properties of matter expected near a QCP. 

Charge ordering (CO) phenomena in quarter-filled layered organic materials 
are observed 
in a wide variety of crystal structures, not limited to 
specific Fermi surface shapes or nesting. This indicates the importance of 
onsite and intersite Coulomb repulsion \cite{Mori,Merino01} 
between $\pi$ electrons as the driving force of CO, and 
in turn implies that electronic correlation effects
similar to those found in half-filled systems are inevitably present.
These should be considered together  with the quantum critical
fluctuations of the order parameter to understand the metallic
properties in the neighborhood of the present Coulomb-driven transition.
The latter can, in principle, differ from more
standard charge density wave instabilities of the Fermi surface. 

In this Letter we analyze theoretically the possible existence of a QCP at a CO
transition driven by the quantum fluctuations associated with strong off-site 
Coulomb repulsion, in the absence of Fermi surface nesting.
The influence of the $T=0$ singular quantum critical point 
on the finite temperature metallic properties is studied based on
finite-$T$ Lanczos  
diagonalization  \cite{Prelovsek,Liebsch} of an
extended Hubbard model on an anisotropic triangular lattice. At quarter
filling ($n=1/2$ hole per molecule), lattice frustration naturally leads to 
charge ordered metallic states with a single CO pattern.
Our main result is the existence of  
a temperature scale $T^*$ in the dynamical and thermodynamic
properties of the system, that
is suppressed as the CO transition is approached from the homogeneous
metal.
The electronic specific heat coefficient 
 at low temperatures is found to be strongly enhanced 
close to the QCP in analogy with studies of quantum criticality in
heavy fermions. 
Following this analogy, non-Fermi liquid behavior
should occur at finite-$T$ in quarter-filled organic conductors 
at the edge of a CO instability and could
be directly probed by several experimentally measurable quantities.
Due to the universal character of the proposed quantum critical
scenario,
our results could also be relevant to broader classes of systems where 
Coulomb-driven CO occurs, not restricted to the particular 
ordering pattern studied here.

\begin{figure}
\epsfig{file=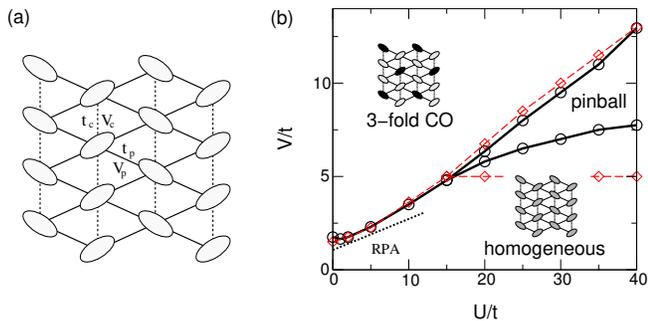,width=8.5cm,angle=0,clip=}
\caption{(color online) 
(a) Lattice structure and microscopic parameters of the 
  extended Hubbard model Eq. (\ref{eq:model}).
(b) Phase diagram obtained at $T=0$ from numerical diagonalization of
  $N_s=12$ (full line) and $18$ sites clusters (dashed).
}
\label{fig:PD}
\end{figure}

\paragraph{Model.}
We focus on
the extended Hubbard model: 
\begin{eqnarray}
H=\sum_{\langle ij\rangle\sigma}t_{ij}(c^\dagger_{i\sigma} c_{j\sigma}+
h.c.) +U \sum_i n_{i\uparrow}n_{i\downarrow}
+\sum_{\langle ij \rangle}V_{ij}n_i n_j
\label{eq:model} 
\end{eqnarray}
on the anisotropic triangular lattice  shown in
Fig. \ref{fig:PD}(a),
where $t_{ij}=(t_p,t_c)$ are the transfer integrals between
nearest neighboring molecules  respectively
along the diagonal ($p$) and vertical ($c$) directions, 
$V_{ij}=(V_p,V_c)$ are the corresponding 
inter-molecular Coulomb interaction
energies and $U$ is the intra-molecular Coulomb repulsion. 
The model Eq. (\ref{eq:model}) has been studied
via a variety of techniques due to its relevance to  $\theta$-type
two-dimensional organic conductors
(see Ref. \cite{Kuroki} for a recent review).
Here we  follow the standard practice and
neglect longer-range Coulomb interactions
\cite{Kuroki} as well as electron-lattice effects \cite{Udagawa}
that are however essential to recover 
the various ordering patterns realized
in these materials.
For the sake of simplicity we consider an isotropic inter-site repulsion
$V_c=V_p\equiv V$ and set  $t_c=0$, $t_p\equiv t>0$.
 This  choice is representative of the 
$\theta$-ET$_2$X salts with X$=$CsCo(SCN)$_4$, X$=$CsZn(SCN)$_4$ and
X$=$I$_3$
where the molecular  orbital overlap is strongly suppressed
along the $c$ direction \cite{Mori,Kuroki}.
These materials lie 
close to 
(on both sides of)
the bandwidth controlled CO transition in Mori's phase diagram \cite{Mori} and 
are therefore optimal candidates for the observation 
of an interplay between critical charge fluctuations 
and electronic correlation effects.

\paragraph{Phase diagram.}

The phase diagram obtained  at $T=0$ from the numerical diagonalization
of the model Eq. (\ref{eq:model}) on $N_s=12$ and $N_s=18$ 
site clusters is presented in Fig. \ref{fig:PD}(b). The different phases can be
identified by analyzing the behavior of the charge correlation
function $N_sC({\bf q})=N_s^{-1}\sum_{ij}\langle n_i n_j \rangle
e^{i{\bf q}\cdot {\bf R}_{ij}}$. In the
thermodynamic limit, this quantity diverges at a single
wavevector ${\bf Q}\neq 0$  at the onset of charge
order. An accurate 
numerical determination of the phase boundaries relying 
on a proper finite-size scaling of the results is
prohibitive for the fermionic system under study, due to the rapidly
increasing size of the Hilbert space. We therefore identify  
the $T=0$ ordering transition, $V_{CO}$, 
as the locus of steepest variation of charge
correlations upon varying the interaction parameters. 
An analogous procedure is used to 
 determine the melting temperature of CO, $T_{CO}$.
In the physically relevant regime explored here, $U/t\lesssim 20$
\cite{footnote}
the phase boundaries agree on the two cluster sizes. 

In the absence of nearest-neighbor repulsion, $V=0$, the system
remains in a homogeneous metallic phase up to arbitrary values of 
the local interaction $U$, as 
holes can effectively avoid each other at concentrations 
away from integer fillings.
An instability towards a charge ordered state with 3-fold periodicity
is realized instead  upon increasing 
the inter-site interaction, $V$, as was
previously obtained by different approaches
\cite{Mori03,Kaneko,Kuroki,Hotta,WatanabeVMC06,NishimotoDMRG08}.
The resulting 3-fold ordering pattern is shown in
Fig. \ref{fig:PD}(b). 

At low and moderate values of
$U/t\lesssim 5$, down to $U=0$, 
the CO transition essentially follows the predictions
of mean-field approaches \cite{Mori03,Kaneko,Kuroki}. A calculation in
the random phase approximation (RPA)  
yields $V_{CO}=1.06t+U/6$, which is shown as a
dotted line in Fig. \ref{fig:PD}(b). This law  is correctly recovered by the
numerical data at low $U$, but sizable deviations appear at as soon as
$U/t\gtrsim 10$ due to the increasing effects
of many-body electronic correlations. The boundary obtained
numerically in this region is independent of the cluster size, and
our value  $V_{CO}/t=3.5$ at $U/t=10$ is in good agreement with 
existing numerical results in larger systems 
\cite{WatanabeVMC06,NishimotoDMRG08}.

Before moving to the analysis of the
correlated metallic phase at the edge of charge order,
let us note that the charge correlation function 
also provides indications of a crossover taking place within the CO phase, separating a
conventional 3-fold state from a more exotic ``pinball liquid''
phase \cite{Hotta}. The latter arises because at large $U$,
mean-field like configurations where charge-poor molecules are completely depleted
become energetically unfavorable, as these imply that each charge rich
molecule should accomodate up to $3/2$ holes on average. To prevent
double occupancy, part of the hole density necessarily spills out and
decouples from the charge rich sublattice, resulting in a 
separate fluid moving freely in the remaining sublattice
\cite{Hotta}.    
This partial ordering, occurring for $V\lesssim U/3$,
 corresponds to a value $C({\bf Q})=n^2/3$,  
to be contrasted with the value $C({\bf Q})=n^2$ obtained in
the 3-fold state  at large $V$. 

\paragraph{The correlated metal close to charge ordering.}

We start by analyzing 
the kinetic energy of the interacting system,
a quantity that provides direct information on how
the motion of the charge carriers is 
hindered by interactions, and can 
be evaluated with good accuracy through finite-$T$ Lanczos diagonalization.
Its importance in correlated systems has been recently recognized
\cite{Millis04,  Qazilbash09},
and resides in the fact that
it can in principle be accessed 
from optical absorption experiments, 
providing a quantitative
measure of many-body correlation effects. 

\begin{figure}
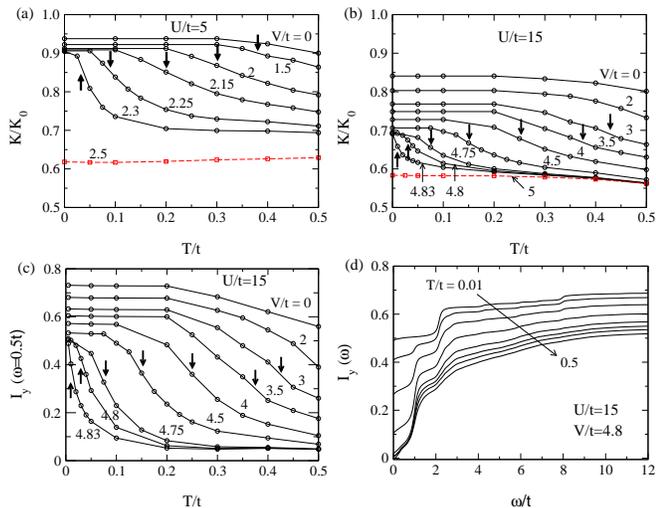

  \centering
\epsfig{file=Fig2_a.eps,width=4.25cm,clip=}
\epsfig{file=Fig2_b.eps,width=4.25cm,clip=}
\epsfig{file=Fig2_c.eps,width=4.25cm,clip=}
\epsfig{file=Fig2_d.eps,width=4.25cm,,clip=}
  \caption{(color online). Kinetic energy of the $N_s=12$ interacting system
  at $U/t=5$ (a) and $U/t=15$ (b); $T$-dependence of low frequency optical weight (c), and integrated optical spectral
    weight $I(\omega)$ for $T/t=0.01,0.03,0.05$ and $0.1-0.5$ in 0.1 intervals  (d). Arrows in (a),(b),(c) correspond to
    inflection points of the curves, defining 
    the temperature scale $T^*$.} 
  \label{fig:kinetic}
\end{figure}

The kinetic energy, $K$, normalized to the non-interacting band value,
$K_0$, is shown in
Figs. \ref{fig:kinetic}(a) and (b), respectively for
$U/t=5$ and $U/t=15$, for several values of  the intersite repulsion
 across the CO transition.
At $U/t=5$ the kinetic energy at $T=0$ 
stays essentially unrenormalized, $ K/ K_0 \gtrsim 0.9$, upon
increasing $V$ all the way up to the CO 
transition occurring at $V_{CO}=2.33t$, as expected in
a weakly correlated Fermi liquid. It then 
suddenly drops to a
value $K/ K_0  \sim 0.6$ upon entering the charge ordered phase. 
This residual value is ascribed to local (incoherent) 
hopping processes in the charge ordered pattern \cite{Millis04} 
and to the motion of 
remnant itinerant electrons not gapped by the 
ordering transition \cite{Kaneko,WatanabeVMC06}.

A richer behavior is revealed by  
the data at finite temperatures, that clearly indicate the
emergence of a temperature scale $T^*$ 
that marks an analogous suppression of the
kinetic energy occurring 
{\it within the homogeneous metallic phase}  (we define $T^*$ as the
locus of steepest variation of $K$ with temperature, denoted by
arrows in Fig. \ref{fig:kinetic}). The scale $T^*$
appears to be entirely controlled by the approach 
to the zero-temperature ordering 
transition, a behavior that is strongly reminiscent of
what is expected close to a QCP.
The situation is similar at $U/t=5$ and  $U/t=15$ [Fig. \ref{fig:kinetic}(b)],
although in the latter case
the kinetic energy ratio at $T=0$ is already reduced down to values 
$K/K_0\sim 0.7$ before entering the CO phase
at $V_{CO}=4.83t$, which is indicative of 
a moderately correlated electron liquid. In this case 
the quantum critical behavior adds up to the correlated electron picture, 
affecting the motion of electrons that have  already been 
slowed down by local electronic correlations.

The above observations can be directly related to the low-energy 
quasiparticle properties by analyzing the temperature dependence of the
integrated optical weight $I(\omega)=\int_0^\omega
\sigma(\omega^\prime) d\omega^\prime$. 
The low-frequency integral 
$I(\omega= 0.5 t)$, reported in  Fig. \ref{fig:kinetic}(c),
comprises most of the quasiparticles contributing to the
Drude behavior in a normal Fermi liquid, while excluding higher-energy
incoherent excitations arising from the strong electronic
interactions.
Comparison of
Figs. \ref{fig:kinetic}(b) and (c)
demonstrates that the strong reduction of kinetic energy  above $T^*$
primarily originates from a drastic suppression 
of the low-energy coherent quasiparticles. 
Finally, Fig. \ref{fig:kinetic}(d) illustrates  $I(\omega)$ at
a given $V/t=4.8$ just below the CO
transition, showing that the quasiparticle weight  lost at $T^*$  is partly 
transferred to high-energy excitations,  that are broadly
distributed on the scale of $U,V$.

\begin{figure}
  \centering
  \epsfig{file=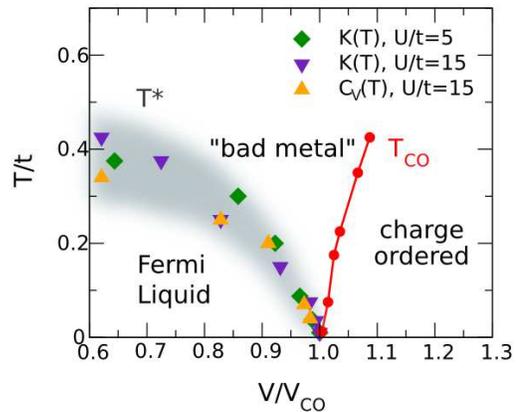,width=7.0cm,angle=0,clip=} 
  \caption{(color online). Finite temperature phase diagram 
illustrating the emergence of non-Fermi liquid properties induced by
the proximity     to  the charge ordering 
transition. The interaction strength has been
    rescaled to the critical values to compare QCP related  properties
    for weakly and moderately correlated systems.
}
  \label{fig:PDT}
\end{figure}

The above results are summarized  in the finite temperature phase
diagram of Fig. \ref{fig:PDT}, that constitutes the central result of
this work.   Scaled units $V/V_{CO}$ are used so that
the weakly ($U/t=5$) and moderately ($U/t=15$) correlated cases 
can be directly compared, illustrating the universality of the $T^*$
phenomenon in proximity to the CO instability. Both $T^*$
and $T_{CO}$ appear to vanish at $V_{CO}$ leading to a funnel-type 
'bad' metallic region with strong quantum critical fluctuations.

\begin{figure}
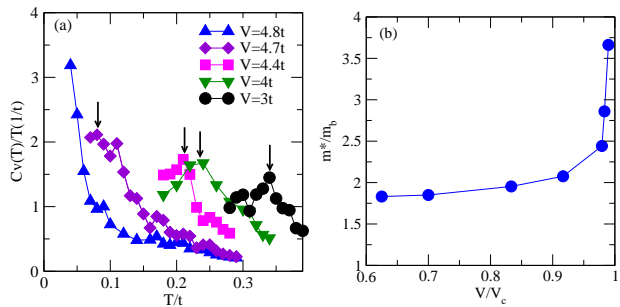

  \centering
  \epsfig{file=CvUV_paper.eps,width=4.cm,angle=0,clip=}
  \epsfig{file=effmassU15.eps,width=4.cm,angle=0,clip=}
  \caption{(color online) 
    (a) Specific heat coefficient $\gamma=C_V/T$ in units
    of $1/t$ for $U/t=15$.
 The curves are shown down to a temperature
    comprising $\sim 10$ excitations in the Lanczos diagonalization.
(b) effective mass ratio, $m^*/m_b$, estimated from the peak
value of $\gamma$ shown in (a).}
  \label{fig:CV-mass}
\end{figure}

To obtain further insight into the behavior of quasiparticles near the
CO instability and make contact with the established concepts of 
quantum criticality,
we have calculated 
the specific heat 
coefficient $\gamma=C_V/T$. 
Our results, reported in 
Fig. \ref{fig:CV-mass}(a), resemble the behavior of 
nearly two-dimensional antiferromagnetic metals in which 
a singular increase 
is expected upon lowering the temperature  close to the QCP, 
crossing over to a constant value at the onset of
Fermi liquid behavior  \cite{Lohneysen,Moriya}. 
Curves similar to those in Fig. \ref{fig:CV-mass}(a) are
commonly observed in heavy fermion systems
\cite{Stewart,Custers,Lohneysen}. 
 As a striking confirmation of the QCP scenario emerging from the preceding
paragraphs, we see that there is a direct correspondence between 
thermodynamic and dynamical properties: the
peak position in the specific heat 
essentially coincides 
with the temperature $T^*$ derived from the kinetic energy  and the
low-frequency optical integral near to the QCP (see  Fig. \ref{fig:PDT}).

Finally, we discuss the behavior of the effective mass 
 as extracted from the low-temperature limit of
the specific heat coefficient (in practice we estimate $m^*$ 
 from the peak value of $\gamma$  to overcome the
numerical limitations of the Lanczos technique at low $T$).
It can be expected on general grounds
that the strong electron-electron interactions responsible for the
3-fold CO will strongly affect those parts of the Fermi
surface that are connected by momenta closest to the ordering
wavevector ${\bf Q}$. This should lead to the emergence of
``hot spots''  with divergent  effective mass,
$m^*/m_b \propto \ln(1/|V-V_{CO}|)$,
in full analogy to the situation encountered in metals
close to a magnetic instability \cite{Moriya,Merino06}.
The effective  mass  reported in
Fig. \ref{fig:CV-mass}(b) indeed shows a marked enhancement 
at the approach of $V_{CO}$, that adds to  a moderate
renomalization $m^*/m_b\lesssim 2$ provided by non-critical electronic
correlations.
Whether the destruction of quasiparticles remains confined to such 'hot
spots' or spreads over the whole Fermi surface is an unsettled issue that  
is also actively debated in the context of heavy fermion 
materials \cite{Coleman,Si}.

\paragraph{Concluding remarks.}

Our results indicate the occurrence of non-Fermi liquid
behavior driven by a combination of electronic
correlations and quantum critical fluctuations
close to a CO instability  in quarter-filled organic conductors. 
Quantum critical behavior has been recently reported in transport
studies of  the quarter-filled compounds $\kappa$-(DHDA-TTP)$_2$SbF$_6$ and
(MeDH-TTP)$_2$AsF$_6$, by tuning the system across the CO transition
via an applied pressure \cite{Yasuzuka,Weng}. 
A stringent verification of our theoretical 
predictions could be achieved in the conductor
$\theta-$(BEDT-TTF)I$_3$, whose band structure is properly described
by the model Eq. (1). Remarkably, this is the only salt of the $\theta$ family
exhibiting superconductivity {\it and} is at the edge of CO. Recent optical studies \cite{Takenaka1} have shown 
a rapid loss of electron coherence
upon increasing the temperature, associated with a marked reduction of
kinetic energy as obtained here, and the evolution of the
integrated spectral weight $I(\omega)$ with temperature found
experimentally  compares very well with our result in
Fig. \ref{fig:kinetic}(d).  
The presence of an unexplained far infrared absorption peak 
whose position is controlled by the
temperature scale  $T$ alone could be a clue
of an emergent collective excitation of the QCP \cite{Caprara}. 
This material undergoes a CO transition under pressure \cite{Tajima},
which could be directly exploited to probe the quantum critical
behavior, applying  the plethora of
experimental techniques that are commonly used in the study of heavy
fermion materials.
Hall coefficient as well as
de Haas-van Alphen experiments appear to be ideal  
probes to test whether quasiparticles are destroyed over the whole Fermi surface 
or on some regions only, shedding light on the nature of the charge order QCP. 

\acknowledgments  L.C. and J.M. acknowledge financial support from
MICINN (CTQ2008-06720-C02-02, CSD2007-00010), and computer
resources and assistance provided by BSC. The authors thank 
I. Paul for useful discussions.

\end{document}